\documentclass[aps,twocolumn,pra,superscriptaddress,amsmath,showpacs,tightenlines]{revtex4-1}
\usepackage{amssymb}
\usepackage{amsmath}
\usepackage{graphicx}
\usepackage{subfigure}
\usepackage{natbib}
\usepackage{epsfig}
\usepackage{amsfonts}
\usepackage{mathrsfs}
\usepackage{CJK}
\usepackage{xcolor}
\usepackage[toc,page,title,titletoc,header]{appendix}

\begin{document}
\title{Driven-dissipative quantum battery with nonequilibrium reservoirs}
\author{Zhihai Wang}
\affiliation{State Key Laboratory of Electroanalytical Chemistry, Changchun Institute of Applied Chemistry, Chinese Academy of Sciences, Changchun 130022, China}
\affiliation{Center for Quantum Sciences and School of Physics, Northeast Normal University, Changchun 130024, China}
\author{Hongwei Yu}
\affiliation{State Key Laboratory of Electroanalytical Chemistry, Changchun Institute of Applied Chemistry, Chinese Academy of Sciences, Changchun 130022, China}
\affiliation{Center for Quantum Sciences and School of Physics, Northeast Normal University, Changchun 130024, China}
\author{Jin Wang}
\email{jin.wang.1@stonybrook.edu}
\affiliation{Department of Chemistry and Department of Physics and Astronomy, State University of New York at Stony Brook, New York 11794, USA}

\begin{abstract}
We investigate a quantum battery system under both external driving and dissipation. The system consists of a coupled two-level charger and battery immersed in nonequilibrium fermionic reservoirs. By considering the changes in the energy spectrum induced by external driving and charger-battery coupling in a non-perturbative manner, we go beyond the secular approximation to derive the Redfield master equation. In the nonequilibrium scenario, both charging efficiency and power of the quantum battery can be optimized through a compensation mechanism. When the charger and battery are off-resonance, a significant chemical potential difference between the reservoirs, which characterizes the degree of nonequilibrium, plays a crucial role. Specifically, the charger's frequency should be higher (lower) than that of the battery when the average chemical potential is negative (positive) to achieve enhanced charging efficiency and power under strong nonequilibrium conditions. Remarkably, the efficiency in the nonequilibrium case can surpass that in the equilibrium setup. Moreover, we find no positive correlation between entanglement and efficiency,  so that the  entanglement is not necessary to enhance the performance of quantum devices. Our results provide insights into the design and optimization of quantum batteries in nonequilibrium open systems.
\end{abstract}
\maketitle
\section{introduction}

\label{sec:intro}

A central challenge in quantum technology is to explore how quantum resources can be utilized to perform tasks that are unattainable with classical systems. One such task arises in the realm of energy storage, giving rise to the concept of the ``quantum battery''. A typical quantum battery system consists of a charger, which supplies energy, and a battery, which stores and releases the energy.

Ever since the concept of the quantum battery was introduced by Alicki and Fannes~\cite{RA2013}, numerous quantum battery models have been proposed in various physical systems. Examples include spin or resonator chain models~\cite{TP2018,DF2018,DR2019,YY2019,JC2020,DR2020,ZL2020,SG2020,VP2020, FC2021,FZ2021,OA2022,FZ2022,YY2022}, Tavis-Cummings and Dicke models in quantum optics~\cite{LF2016,GM2019,DZ2020,AC2020s,WL2021,FQ2022}, and Rydberg atom systems~\cite{YY2021}. In some of these systems, Floquet engineering has been employed to enhance the performance of quantum batteries~\cite{SY2020,SY2021,SM2022}. Given that quantum systems are inherently coupled to their environments, quantum batteries in open systems are now drawing increasing attention~\cite{FB2019,CL2019m,DF2019i,AC2020,MC2020x,FH2020,FT2020,FT,WC2021,KX2021,FM2022}.

{The simplest open system for a quantum battery is perhaps a setup involving two coupled two-level systems, each interacting with external environments. One of these systems, driven by a classical field, acts as the charger, while the other serves as the battery~\cite{DF2019i,shao2025}. To explore how particle exchange between the system and reservoirs impacts the quantum battery's performance, we further couple the charger and battery to two independent fermionic reservoirs. The nonequilibrium effect is characterized by the chemical potential difference between these reservoirs. In this quantum open system, the external driving applied to the charger provides energy to the system, necessitating a reassessment of efficiency. Moreover, the nonequilibrium nature of the reservoirs plays a crucial role in shaping the steady-state properties of the system.}

{To analyze the effects of driving and nonequilibrium effects on the performance of the quantum battery, we employ a quantum master equation approach under the Born-Markovian approximation~\cite{HP2007}. Classical driving significantly alters the energy spectrum of the charger-battery system. However, in the conventional Lindblad master equation approach, driving is treated as a perturbation, its impact on the system's energy spectrum is neglected, and the injected energy is not fully considered when evaluating the efficiency of the quantum battery. To address these limitations, we go beyond the Lindblad master equation and adopt the Redfield master equation~\cite{FB1957,AG1957}, which has been extensively used in studies of quantum transport~\cite{AI,CK2015} and photosynthetic processes~\cite{VI2004,JJ2015,MJ2020}. Specifically, we work in the eigen state representation by considering the case of strong coupling between the driving field and the system, thus treating the driving field as an integral part of the open system rather than as a perturbation. This approach allows us to define the charging efficiency and power in a consistent manner. This methodology contrasts sharply with the traditional treatment~\cite{DF2019i}, where the driving is modeled phenomenologically as an effective reservoir weakly coupled to the system. Additionally, we consider the nonequilibrium effects characterized by the chemical potential difference between the two reservoirs coupled to the system. Such nonequilibrium has been shown to induce steady-state coherence and entanglement~\cite{ZD2014,SW2015,ZD2015,ZD12015,YH2018,ZW2018,GG2018,ZW2019,LB2020,BM2021,YK2022}. By moving beyond the secular approximation, we investigate whether the nonequilibrium reservoirs can enhance the performance of the quantum battery.}

Using the Redfield master equation, we uncover a compensation mechanism that enhances the charging efficiency and power of the quantum battery. Specifically, the charger's frequency should exceed that of the battery when the average chemical potential is sufficiently negative and large. Under these conditions, the maximum efficiency can exceed 90\%\, significantly surpassing the efficiency achievable in the equilibrium case. Moreover, this compensation mechanism is also effective in boosting the charging power of the quantum battery. {In addition, it has long been believed that entanglement enhances the performance of quantum devices. Our findings challenge this notion,  as demonstrated with a model-independent theory~\cite{JY2024}, we observe no positive correlation between charger-battery entanglement and the efficiency of the quantum battery.}

The rest of the paper is organized as follows. In Sec.~\ref{model}, we illustrate our model and derive the Redfield master equation. In Sec.~\ref{steady}, we discuss the steady state entanglement in our system. The charging efficiency and power of quantum battery are investigated in Sec.~\ref{efficiency1} and Sec.~\ref{power1}, respectively. At last, we give a short conclusion in Sec.~\ref{con}.

\section{Model and master equation}
\label{model}

\begin{figure}
\centering
\includegraphics[width=0.9\columnwidth]{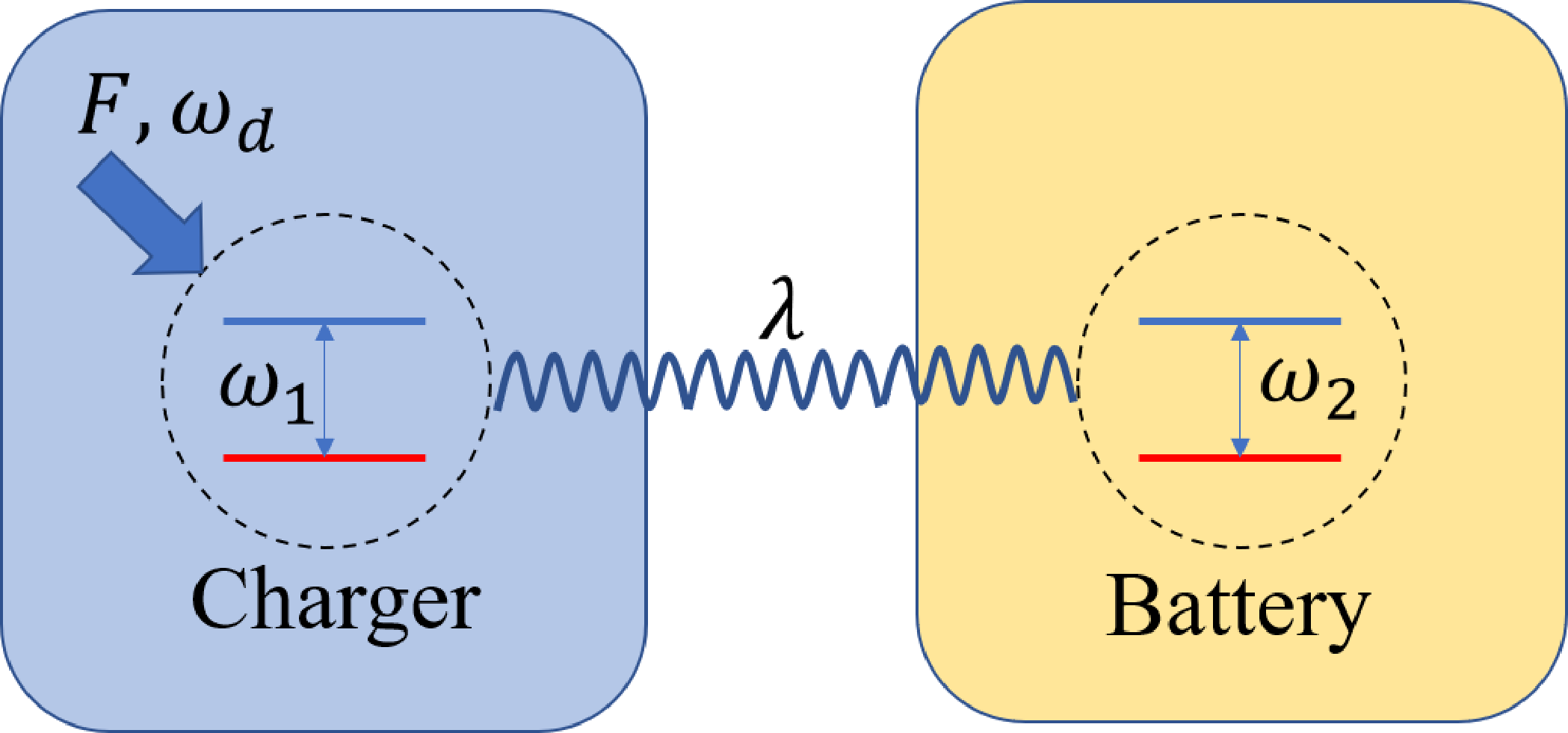}
\caption{Schematic diagram of the quantum battery model under consideration. The charger and the battery are modeled as two-level systems with transition frequencies $\omega_1$ and $\omega_2$, respectively. Each system is coupled to its individual reservoir and they interact with each other via a coupling strength $\lambda$. The charger is driven by an external classical field with driving strength $F$ and frequency $\omega_d$.
}
\label{scheme}
\end{figure}

As schematically illustrated in Fig.~\ref{scheme}, the quantum battery system under consideration consists of a charger and a battery, both modeled as two-level systems. The charger is driven by a classical field, and both the charger and the battery are coupled to their respective reservoirs, which obey fermionic statistics. In the rotating frame with respect to the driving field frequency $\omega_d$, the Hamiltonian of the entire system, including the reservoirs, is given by $H=H_s+H_B+V$.

The Hamiltonian for the charger-battery system is~\cite{DF2019i} ($\hbar=k_B=1$ in what follows)

\begin{eqnarray}
H_{s}&=&\frac{\omega_1}{2}\sigma_{z}^{(1)}+\frac{\omega_2}{2}\sigma_{z}^{(2)}
+\lambda\left(\sigma_{+}^{(1)}\sigma_{-}^{(2)}+\sigma_{-}^{(1)}\sigma_{+}^{(2)}\right) \nonumber\\&&
+\frac{F}{2}\left(\sigma_{+}^{(1)}e^{-i\omega_d t}+\sigma_{-}^{(1)}e^{i\omega_d t}\right).
\label{H1}
\end{eqnarray}

Here, $\sigma_m^{(i)} (m = z, +, -)$ represents the Pauli operators for the $i$th two-level system with transition frequency $\omega_i$. The indices $1$ and $2$ correspond to the charger and the battery, respectively. The driving field has a frequency $\omega_d$, $F$ denotes the driving strength, and $\lambda$ represents the coupling strength between the charger and the battery.

The Hamiltonian of the reservoirs is given by
\begin{equation}
H_{B} = \sum_{k}\omega_{bk}b_{k}^{\dagger}b_{k} + \sum_{k}\omega_{ck}c_{k}^{\dagger}c_{k},
\end{equation}
where $b_k$ ($b^{\dagger}_k$) and $c_k$ ($c^{\dagger}_k$) are the annihilation (creation) operators for the $k$th mode with frequencies $\omega_{bk}$ and $\omega_{ck}$ in the reservoirs coupled to the charger and the battery, respectively.

{The Hamiltonian describing the system-reservoir coupling is expressed as
\begin{eqnarray}
V & = & \sum_{k}g_{k}\left(\sigma_{-}^{(1)}b_{k}^{\dagger} + \sigma_{+}^{(1)}b_{k}\right)
+ \sum_{k}f_{k}\left(\sigma_{-}^{(2)}c_{k}^{\dagger} + \sigma_{+}^{(2)}c_{k}\right),\nonumber \\
\label{inter}
\end{eqnarray}
where $g_k$ ($f_k$) denotes the coupling strength between the charger (battery) and the $k$th mode in its respective reservoir.}

{The time dependence of the Hamiltonian can be eliminated by working in the rotating frame defined by $U_1(t) = \exp\left[i\omega_d\left(\frac{\sigma_z^{(1)} + \sigma_z^{(2)}}{2} + \sum_k (b_k^{\dagger}b_k + c_k^{\dagger}c_k)\right)t\right]$.}

In this rotating frame, the Hamiltonian becomes

\begin{eqnarray}
H_{s} &=& \frac{\Delta_1}{2}\sigma_{z}^{(1)} + \frac{\Delta_2}{2}\sigma_{z}^{(2)}
+ \lambda\left(\sigma_{+}^{(1)}\sigma_{-}^{(2)} + \sigma_{-}^{(1)}\sigma_{+}^{(2)}\right)\nonumber \\
&&+ \frac{F}{2}\left(\sigma_{+}^{(1)} + \sigma_{-}^{(1)}\right), \label{rotating} \\
H_{B} &=& \sum_{k}(\omega_{bk} - \omega_d)b_{k}^{\dagger}b_{k}
+ \sum_{k}(\omega_{ck} - \omega_d)c_{k}^{\dagger}c_{k},
\end{eqnarray}
where $\Delta_i = \omega_i - \omega_d$, and the interaction Hamiltonian $V$ remains unchanged due to the application of the rotating wave approximation.

With these significant approximations and simplifications, the conventional Markovian master equation for the charger-battery system can be written as
\begin{eqnarray}
\frac{d}{dt}\rho &=& -i[H_s, \rho]
+ \sum_{i=1}^{2} J_i(\omega_i)N_i(\omega_i)D_{\sigma_+^{(i)}}[\rho]\nonumber \\&&
+ \sum_{i=1}^{2} J_i(\omega_i)\mathcal{N}_i(\omega_i)D_{\sigma_-^{(i)}}[\rho], \label{master0}
\end{eqnarray}
where $D_A[\rho] = 2A\rho A^\dagger - A^\dagger A\rho - \rho A^\dagger A,$
$J_1(\omega) = \pi\sum_k g_k^2\delta(\omega - \omega_{bk})$ and
$J_2(\omega) = \pi\sum_k f_k^2\delta(\omega - \omega_{ck})$ are the spectral densities of the two reservoirs.

For the fermionic reservoirs, the average particle number at frequency $\omega$ in the $i$th reservoir is $N_i(\omega) = \frac{1}{\exp[(\omega - \mu_i)/T_i] + 1}$, where $\mu_i$ and $T_i$ are the chemical potential and temperature of the $i$th reservoir, respectively, and $\mathcal{N}_i(\omega) = 1 - N_i(\omega)$.

Actually, in writing the above  master equation, one only simply adds the dissipation terms of the charger and the battery. As a result, the master equation yields a Lindblad form. We should note that both of the driving to the charger and the charger-battery coupling will affect the eigen spectrum of the system, compared to the case when both of the charger and the battery are free. In what follows, we will derive the master equations by taking into account of both of the strong driving and coupling effects and obtain the Redfield master equations.

\subsection{Redfield master equation}

To obtain the Redfield master equation, we first solve the eigen energies and corresponding eigen states of the Hamiltonian given by Eq.~(\ref{rotating}). This is achieved numerically, and the results are expressed as
\begin{equation}
H_s = \sum_{i=1}^{4}E_i|E_i\rangle\langle E_i|,
\end{equation}
where the eigen energies are ordered as $E_1 > E_2 > E_3 > E_4$, and $|E_i\rangle$ represents the corresponding eigen state.

In terms of the eigen states, we have
\begin{equation}
\sigma_{-}^{(m)} = \sum_{i,j=1}^{4}\chi_{ij}^{(m)}\tau_{ij},
\end{equation}
{where $\tau_{ij} = |E_i\rangle \langle E_j|$, and $\chi_{ij}^{(m)} = \langle E_i| U \sigma_-^{(m)} U^\dagger |E_j\rangle$ for $m=1,2$. The unitary transformation $U$, obtained via numerical diagonalization of the Hamiltonian $H_s$, connects the eigen states with the bare states.}

The relationships between the eigen states and bare states are given by
\begin{subequations}
\begin{eqnarray}
(|E_1\rangle, |E_2\rangle, |E_3\rangle, |E_4\rangle)^T &=& U (|ee\rangle, |eg\rangle, |ge\rangle, |gg\rangle)^T, \\
(|ee\rangle, |eg\rangle, |ge\rangle, |gg\rangle)^T &=& U^{\dagger} (|E_1\rangle, |E_2\rangle, |E_3\rangle, |E_4\rangle)^T.
\end{eqnarray}
\end{subequations}
{By redefining the bare basis of the charger-battery system as $|1\rangle = |ee\rangle$, $|2\rangle = |eg\rangle$, $|3\rangle = |ge\rangle$, and $|4\rangle = |gg\rangle$, the elements of the transformation matrix $U$ are expressed as $U_{ij} = \langle E_i | j \rangle$.}

As a result, the interaction Hamiltonian in the interaction picture can be written as
\begin{eqnarray}
V &=& \sum_{i,j>i}\sum_{k}\left(g_{k}\chi_{ij}^{(1)}b_{k}\tau_{ij}
e^{-i(\omega_{bk}-\omega_{ji})t}+{\rm H.c.}\right)\nonumber \\
&+&\sum_{i,j>i}\sum_{k}\left(f_{k}\chi_{ij}^{(2)}c_{k}\tau_{ij}e^{-i(\omega_{ck}-\omega_{ji})t} + {\rm H.c.}\right),
\end{eqnarray}
where we have applied the rotating wave approximation again, considering that the system weakly couples to the reservoirs.

Using the general form of the master equation~\cite{HP2007},
\begin{equation}
\frac{d}{dt}\rho^{(I)} = -\int_{0}^{\infty}d\tau \, {\rm Tr}_{B}\left[V_{I}(t), \left[V_{I}(t-\tau), \rho^{(I)} \otimes \rho_{B}^{(I)}\right]\right],
\end{equation}
where the superscript $I$ indicates the interaction picture (while the absence of a superscript refers to the Schr\"{o}dinger picture), the master equation under the Markovian approximation for our system reads
\begin{equation}
\frac{d}{dt}\rho = \mathcal{L}\rho = -i\left[\sum_{j=1}^{4}E_j|E_j\rangle\langle E_j|, \rho\right] + \mathcal{D}_1(\rho) + \mathcal{D}_2(\rho),
\label{master3}
\end{equation}
where
\begin{eqnarray}
\mathcal{D}_1(\rho) &=& \sum_{i,j>i}\sum_{m,n>m}\sum_{\alpha=1,2}J_{\alpha}
(\epsilon_{mn})\mathcal{N}_{\alpha}(\epsilon_{mn})
\mathcal{G}^{(\alpha)}_{ij,mn}[\rho], \nonumber \\ \\
\mathcal{D}_2(\rho) &=& \sum_{i,j>i}\sum_{m,n>m}\sum_{\alpha=1,2}J_{\alpha}
(\epsilon_{mn})N_{\alpha}(\epsilon_{mn})\mathcal{H}^{(\alpha)}_{ij,mn}[\rho],\nonumber \\
\end{eqnarray}
and
\begin{eqnarray}
\mathcal{G}^{(\alpha)}_{ij,mn}[\rho] &=& \chi_{ji}^{(\alpha)}\chi_{mn}^{(\alpha)}\left(\tau_{ji}\rho\tau_{mn}
- \rho\tau_{mn}\tau_{ji}\right)\nonumber \\&&
+ \chi_{ij}^{(\alpha)}\chi_{nm}^{(\alpha)}\left(\tau_{nm}\rho\tau_{ij}
- \tau_{ij}\tau_{nm}\rho\right), \\
\mathcal{H}^{(\alpha)}_{ij,mn}[\rho] &=& \chi_{ij}^{(\alpha)}\chi_{nm}^{(\alpha)}\left(\tau_{ij}\rho\tau_{nm}
- \rho\tau_{nm}\tau_{ij}\right)\nonumber \\&&
+ \chi_{ji}^{(\alpha)}\chi_{mn}^{(\alpha)}\left(\tau_{mn}\rho\tau_{ji}
- \tau_{ji}\tau_{mn}\rho\right).
\end{eqnarray}

{One should note that, since the charger and the battery are immersed in fermionic reservoirs characterized by different chemical potentials which creates nonequilibrium conditions, we have gone beyond the commonly used secular approximation in the above Redfield master equation. This is achieved by including the summation terms with $\epsilon_{mn} \neq \epsilon_{ij}$, which correspond to fast oscillating terms in the interaction picture. In our previous works~\cite{ZW2018,ZW2019}, we demonstrated that these non-secular terms give rise to steady-state coherence, a feature that is absent in equilibrium open systems.}

\subsection{Basic concept for quantum battery}

In the eigen state representation, the initial state is given by $|\psi(0)\rangle_e = U^{\dagger}|\psi(0)\rangle$, and the system undergoes time evolution governed by the master equation in Eq.~(\ref{master3}). After a time interval $\tau$, at which the charging process is assumed to end, the system reaches a state with density matrix $\rho(\tau)$. Returning to the bare representation and the Schr\"{o}dinger picture, we have
\begin{equation}
\tilde{\rho}(\tau) = U_1(\tau) U \rho(\tau) U^{\dagger} U_1^{\dagger}(\tau),
\end{equation}
{where $U_1(\tau) = \exp[i\omega_d(\sigma_z^{(1)}+\sigma_z^{(2)})\tau/2]$ is defined by the rotating frame, and $U$ is the unitary transformation connecting the bare state representation and the eigen state representation.}

We formally assume that $\tilde{\rho}(\tau)$ can be expressed as (in the basis of $\{|ee\rangle, |eg\rangle, |ge\rangle, |gg\rangle\}$)
\begin{equation}
\tilde{\rho}(\tau) = \left(\begin{array}{cccc}
M_{11} & M_{12} & M_{13} & M_{14} \\
M_{21} & M_{22} & M_{23} & M_{24} \\
M_{31} & M_{32} & M_{33} & M_{34} \\
M_{41} & M_{42} & M_{43} & M_{44}
\end{array}\right),
\end{equation}
and the reduced density matrix for the battery subsystem can be expressed as (in the basis of $\{|e\rangle, |g\rangle\}$)
\begin{equation}
\rho_{B} = \left(\begin{array}{cc}
M_{33} + M_{11} & M_{12} + M_{34} \\
M_{21} + M_{43} & M_{44} + M_{22}
\end{array}\right).
\end{equation}

In the quantum battery scenario, one of the key quantities of interest is the mean charging energy $E_B$ stored in the battery at the end of the charging process, which is expressed as~\cite{AE2004,WP1978}
\begin{eqnarray}
E_B(\tau) &=& {\rm Tr}[H_B \rho_B(\tau)] - {\rm Tr}[H_B \rho_B(0)], \label{EB}
\end{eqnarray}
where $H_B = \omega(|e\rangle\langle e| - |g\rangle\langle g|)/2$. The corresponding average energy per unit time, i.e., the charging power, is given by $P(\tau) = E_B(\tau)/\tau$. Another quantity of interest is the ergotropy function $\mathcal{E}_B$, which has a clear physical interpretation as follows.

Considering the state of a quantum system, characterized by the free Hamiltonian $H$, is given by the density matrix $\rho$. Using spectral decomposition, $\rho$ and $H$ can be expressed as
\begin{eqnarray}
\rho = \sum_{n}r_n|r_n\rangle\langle r_n|, \quad H = \sum_{n}e_n|e_n\rangle\langle e_n|,
\end{eqnarray}
where $r_n$ and $e_n$ are the eigen values of $\rho$ and $H$, with corresponding eigen states $|r_n\rangle$ and $|e_n\rangle$. By arranging the eigen values such that $r_0 \geq r_1 \geq r_2 \geq \cdots$ and $e_0 \leq e_1 \leq e_2 \leq \cdots$, we can construct a quantum state with a density matrix given by
\begin{equation}
\rho^{(p)} = \sum_n r_n|e_n\rangle \langle e_n|.
\end{equation}
When a quantum system is in such a state, it cannot release energy to its surroundings. Therefore, the state $\rho^{(p)}$ is referred to as a passive state~\cite{AE2004,WP1978}. The energy of the passive state is given by
\begin{equation}
E^{(p)} = {\rm Tr}(H\rho^{(p)}) = \sum_{n}r_n e_n,
\end{equation}
which can also be expressed as
\begin{equation}
E^{(p)} = \min_{U_B}{\rm Tr}[H U_B \rho_{B}(\tau) U_B^{\dagger}].
\end{equation}
Here, the minimization is performed over all local unitary transformations $U_B$ acting on the battery subsystem. This represents the portion of energy in $E_B(\tau)$ that is locked in correlations within the system, making it inaccessible through local operations on the battery.

\begin{figure*}
\centering
\includegraphics[width=0.66\columnwidth]{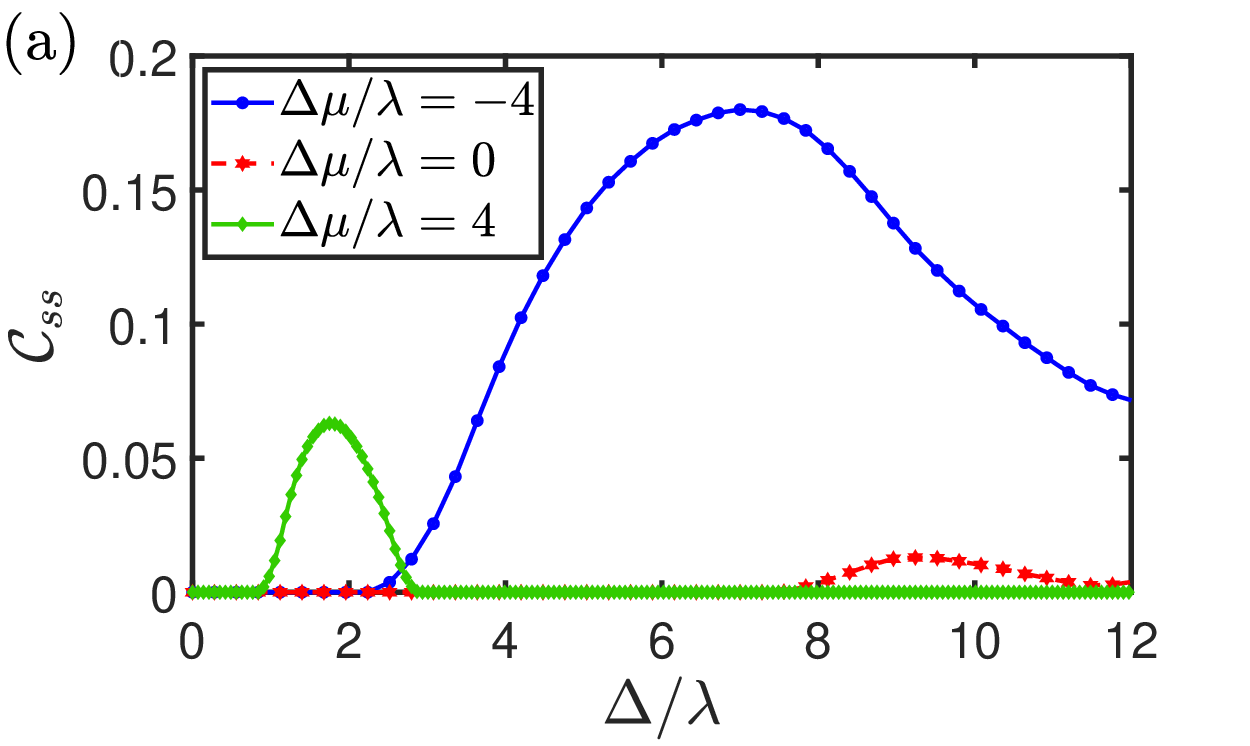}
\includegraphics[width=0.66\columnwidth]{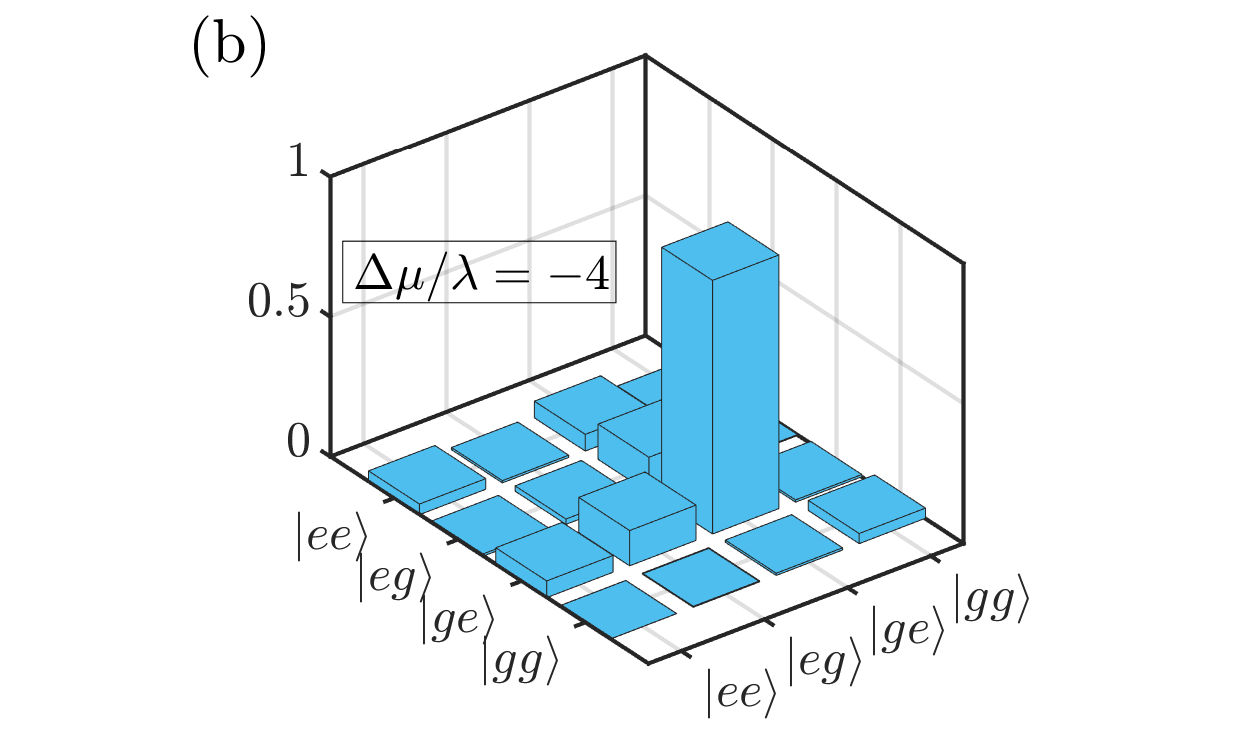}
\includegraphics[width=0.66\columnwidth]{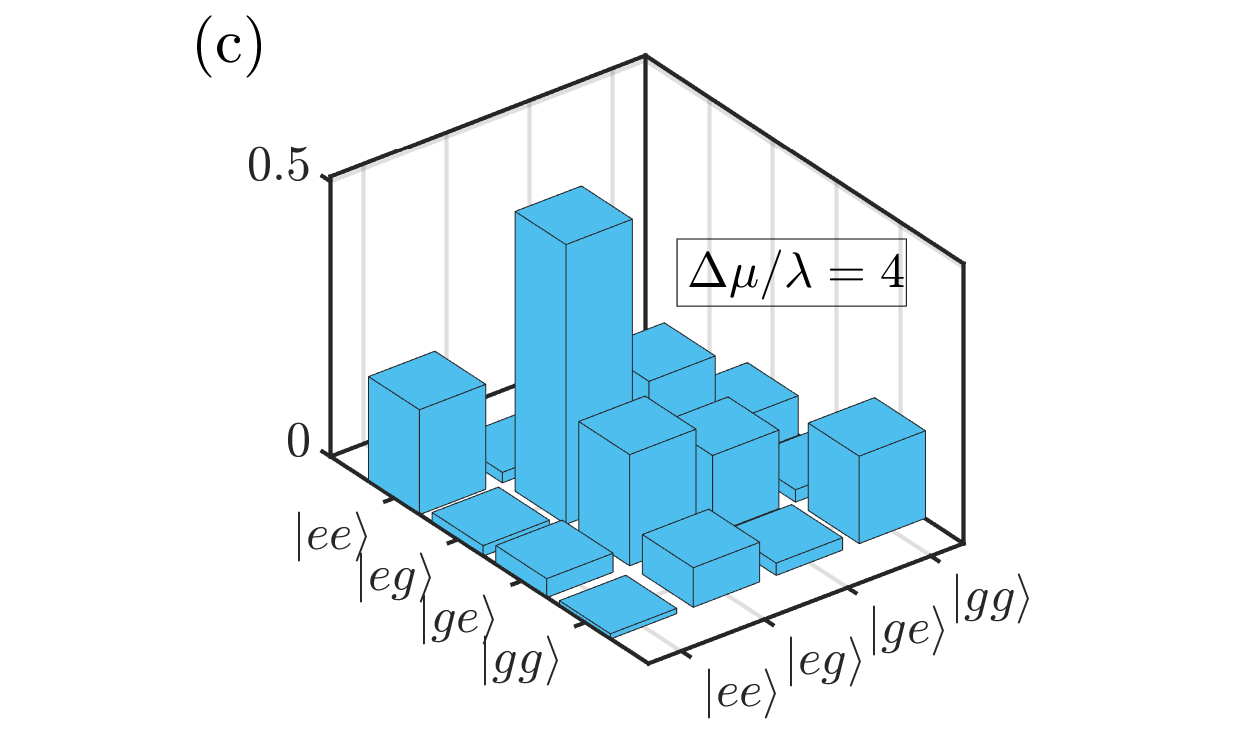}
\caption{(a) The concurrence of the steady states governed by the Redfield master equation under fermion reservoirs. (b) and (c) Tomography of the steady states under nonequilibrium fermion reservoirs, where $\Delta$ is chosen such that the corresponding concurrence reaches its maximum value. The parameters are set as $F = 0.5\lambda$, $\omega_c = 5\lambda$, $T_1 = T_2 = \lambda$, $\bar{\mu} = 2\lambda$, and $\alpha_1 = \alpha_2 = 0.1\lambda$.}
\label{concurrence}
\end{figure*}

Thus, the ergotropy, which represents the extractable energy from the battery, is defined as
\begin{eqnarray}
\mathcal{E}_{B}(\tau) = E_{B}(\tau) - E^{(p)}. \label{ergb}
\end{eqnarray}

For our two-level battery system, the above quantities can be calculated as~\cite{DF2019i}
\begin{eqnarray}
E_B(\tau) &=& \frac{\omega}{2}(N+1), \\
\mathcal{E}_{B}(\tau) &=& \frac{\omega}{2}\left(\sqrt{\langle\sigma_{z}\rangle^2
+ 4\langle\sigma_{+}\rangle\langle\sigma_{-}\rangle}
+ \langle\sigma_{z}\rangle\right) \nonumber \\
&=& \frac{\omega}{2}\left(\sqrt{N^2 + 4|\mathcal{N}|^2} + N\right),
\end{eqnarray}
where $N = M_{11} + M_{33} - M_{22} - M_{44}$ and $\mathcal{N} = M_{12} + M_{34}$.

In what follows, we will first demonstrate the entanglement between the charger and the battery. Next, we will discuss the charging efficiency $R(\tau) = \mathcal{E}_B(\tau) / E_B(\tau)$ and the average power $P(\tau) = E_B(\tau) / \tau$.

\section{Steady state entanglement}
\label{steady}
Under the combined effects of coherent driving by the classical field and dissipation induced by two non-equilibrium reservoirs, the charger-battery coupled system eventually reaches a steady state. This steady state is generally a mixed state but exhibits a certain degree of entanglement.

{In the pioneering work of Hill and Wootters~\cite{WK1998}, the entanglement of an arbitrary two-qubit state $\rho$ is quantified by the so-called concurrence. The concurrence is defined as $\mathcal{C} = \max\{0, v_1 - v_2 - v_3 - v_4\}$, where $v_1 > v_2 > v_3 > v_4$ are the eigen values of the matrix $\tilde{\rho} = (\sigma_y \otimes \sigma_y) \rho^* (\sigma_y \otimes \sigma_y)$,
evaluated in the basis $\{|ee\rangle, |eg\rangle, |ge\rangle, |gg\rangle\}$. The concurrence $\mathcal{C}$ takes values between $0$ and $1$: the state is separable if $\mathcal{C} = 0$, while it is maximally entangled if $\mathcal{C} = 1$.}

Without loss of generality, we choose the Ohmic spectral density given by
\begin{equation}
J_i(\omega) = \alpha_i \omega \exp(-\omega / \omega_c),
\label{ohmic}
\end{equation}
for $i = 1, 2$. In Fig.~\ref{concurrence}(a), we plot the steady-state concurrence $\mathcal{C}_{\rm ss}$ between the charger and the battery as a function of their detuning $\Delta = \Delta_1 - \Delta_2$, considering both equilibrium and non-equilibrium scenarios where the two reservoirs share the same temperature. To evaluate any physical quantity $A$ at steady state, denoted as $A_{\rm ss} = A(\tau = \infty)$, we set $\tau = 20000 / \lambda$. We have verified that the system achieves its steady state within this time interval.

For the equilibrium case of $\Delta\mu = 0$, the concurrence remains consistently very small ($\mathcal{C}_{\rm ss} < 0.05$) as shown in Fig.~\ref{concurrence}(a), indicating that the charger and the battery are nearly separable. In contrast, non-zero entanglement can be observed when the frequency of the charger exceeds that of the battery for both $\Delta\mu > 0$ and $\Delta\mu < 0$. This suggests that driving the charger at a higher frequency facilitates the generation of steady-state entanglement. Moreover, the maximum steady-state concurrence achieved for $\Delta\mu < 0$ surpasses that for $\Delta\mu > 0$.

In the nonequilibrium case, the concurrence exhibits a non-monotonic behavior as a function of $\Delta$. In particular, we tomograph the steady state when the concurrence reaches its maximum value for $\Delta\mu / \lambda = \pm 4$, as shown in Figs.~\ref{concurrence}(b) and (c). For the negative chemical potential difference, depicted in Fig.~\ref{concurrence}(b), the steady state predominantly occupies the superposition states of $|eg\rangle$ and $|ge\rangle$. Conversely, in Fig.~\ref{concurrence}(c), the mixing of the $|ee\rangle$ and $|gg\rangle$ states for $\Delta\mu > 0$ suppresses the concurrence of the charger-battery system.

\section{Efficiency of the quantum battery}
\label{efficiency1}

In this section, we will discuss the efficiency $R_{\rm ss}$ of the quantum battery setup by considering that the reservoirs are characterized by the Ohmic spectrum in Eq.~(\ref{ohmic}).

Firstly, we consider the case where the charger and the battery are resonant, and the charger is resonantly driven. Furthermore, the charger and the battery are immersed in equilibrium reservoirs, characterized by $T_1 = T_2 = T$, $\mu_1 = \mu_2 = \mu$, and $\Delta_1 = \Delta_2 = 0$.

\begin{figure}
\centering
\includegraphics[width=0.48\columnwidth]{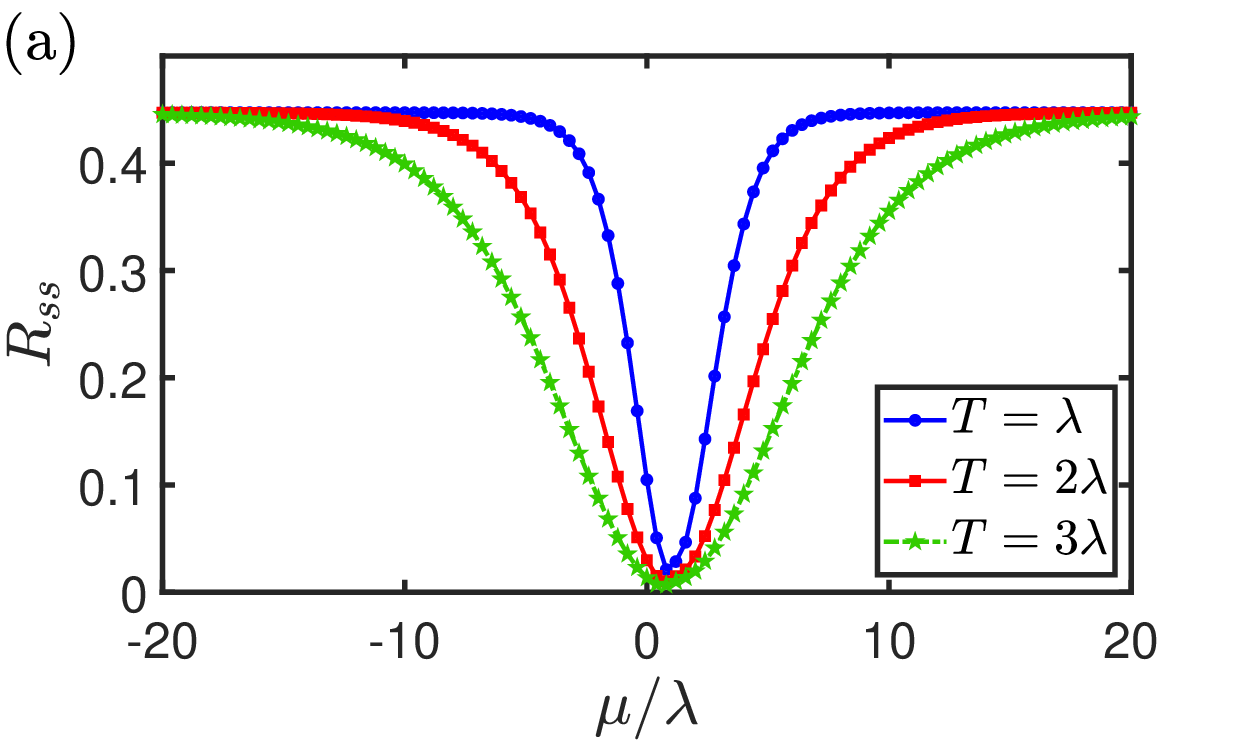}
\includegraphics[width=0.48\columnwidth]{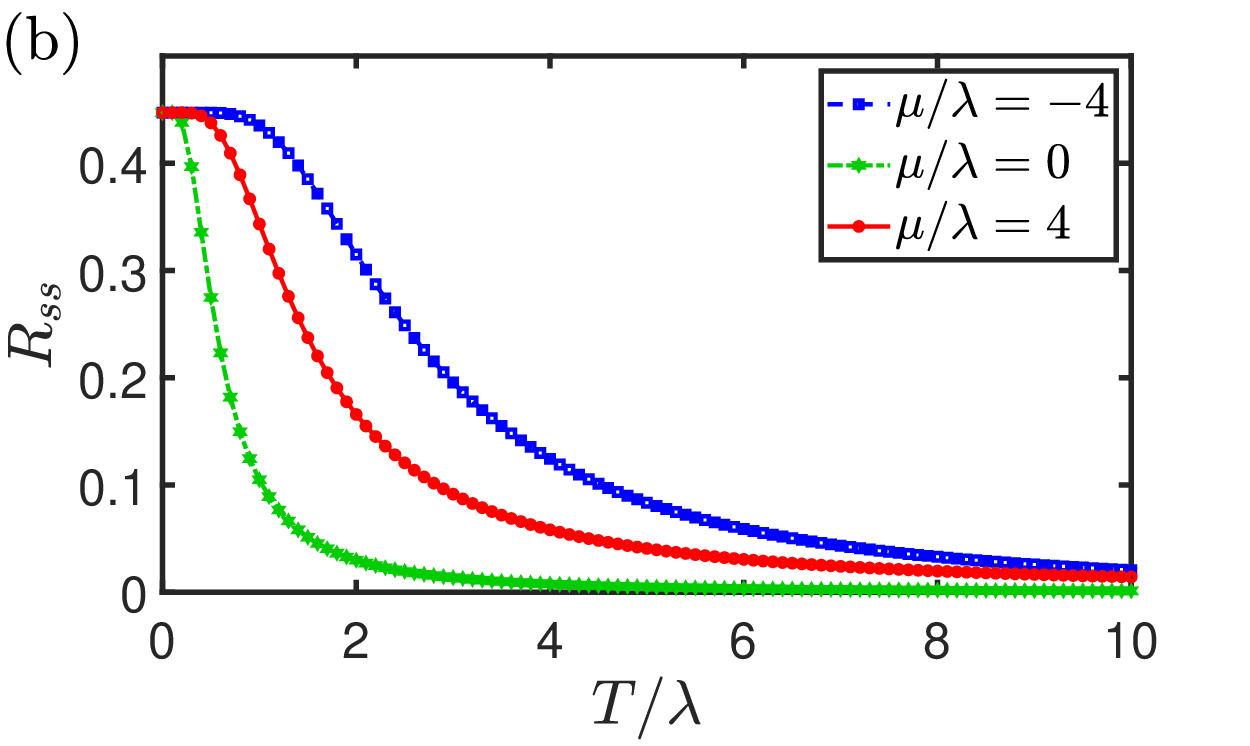}
\caption{The efficiency $R_{\rm ss}$ as a function of $\mu_1 = \mu_2 = \mu$ (a) and $T_1 = T_2 = T$ (b) in equilibrium fermion reservoirs. The parameters are set as $F = 0.5\lambda$, $\omega_c = 5\lambda$, $\Delta_1 = \Delta_2 = 0$, and $\alpha_1 = \alpha_2 = 0.1\lambda$.}
\label{equilibrium_efficiency}
\end{figure}

{In Fig.~\ref{equilibrium_efficiency}(a), we plot the steady-state efficiency as a function of the chemical potential $\mu$ for different temperatures $T$. The results indicate that the efficiency drops off at a specific value of $\mu$, which is slightly greater than $0$ and independent of $T$. A possible explanation for this behavior is that when the chemical potentials of reservoirs are both near zero, they cannot exchange the particles with the charger-battery system, resulting in a low (or possibly zero) efficiency. The slight deviation observed may originate from the external driving of the charger. Additionally, as $\mu$ becomes sufficiently positive or negative, the efficiency saturates at a relatively large value, regardless of the temperature.}

In Fig.~\ref{equilibrium_efficiency}(b), we investigate the efficiency as a function of temperature for different chemical potentials. At very low temperatures, the efficiency is nearly independent of $T$. However, as the temperature increases, the efficiency drops sharply and eventually approaches zero. This indicates that high temperatures degrade the performance of the battery due to thermal fluctuations.

\begin{figure}
\centering
\includegraphics[width=1\columnwidth]{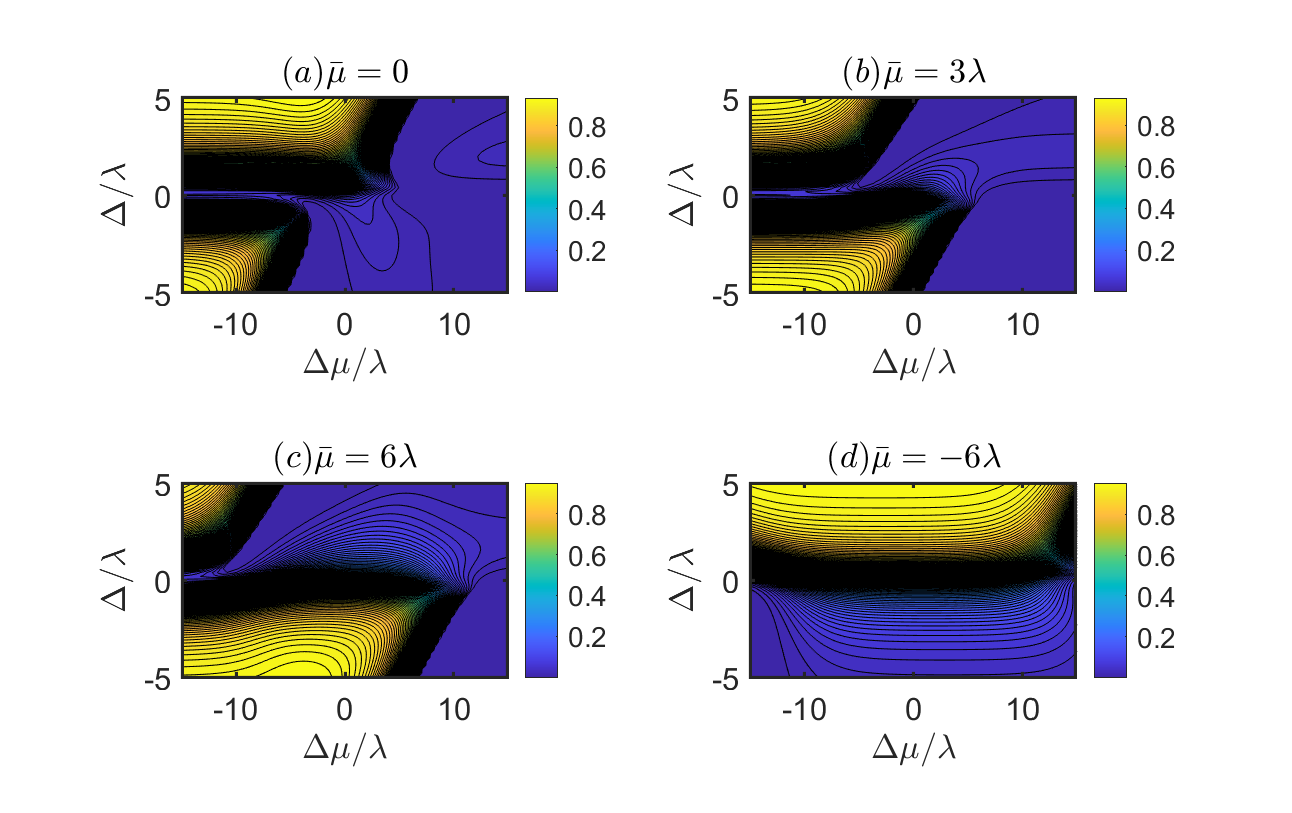}
\caption{The efficiency $R_{\rm ss}$ as a function of the chemical potential difference $\Delta\mu$ and the detuning $\Delta$ for nonequilibrium fermion reservoirs. The parameters are set as $F = 0.5\lambda$, $\omega_{c} = 5\lambda$, $T_1 = T_2 = \lambda$, $\bar{\Delta} = 0$, and $\alpha_1 = \alpha_2 = 0.1\lambda$.
}
\label{nonequilibrium_efficiency}
\end{figure}

Next, we consider more general cases where the charger and the battery are not resonant, and the two reservoirs are not in equilibrium. In Fig.~\ref{nonequilibrium_efficiency}, we plot the efficiency $R_{\rm ss}$ as a function of the chemical potential difference and the charger-battery detuning for different average chemical potentials $\bar{\mu} = (\mu_1 + \mu_2)/2$.

For small $\bar{\mu}$, such as $\bar{\mu} = 0$ and $\bar{\mu} = 3\lambda$, the maximum efficiency is achieved in the parameter regime with either $\Delta > 0$ or $\Delta < 0$ ($\Delta = \Delta_1 - \Delta_2$, the frequency difference between the charger and the battery) under significant chemical potential difference or nonequilibrium, as shown in Figs.~\ref{nonequilibrium_efficiency}(a) and (b). For a positively large $\bar{\mu} = 6\lambda$, the optimal regime shifts primarily to $\Delta < 0$, as depicted in Fig.~\ref{nonequilibrium_efficiency}(c). In stark contrast, when $\bar{\mu} = -6\lambda$, the optimal regime transitions to $\Delta > 0$ as seen in Fig.~\ref{nonequilibrium_efficiency}(d).

These behaviors can be intuitively explained by the compensation mechanism. When the chemical potential is negative, the charger-battery system tends to release particles to the environment. To compensate for this particle loss and achieve higher efficiency, the charger frequency should be higher than that of the battery (\(\Delta > 0\)). Conversely, when the chemical potential is positive, the charger frequency should be lower than the battery (\(\Delta < 0\)). Thus, the highest efficiency is achieved in the regime $\Delta > 0$ for $\bar{\mu} < 0$ (Fig.~\ref{nonequilibrium_efficiency}(d)) and in the regime $\Delta < 0$ for $\bar{\mu} > 0$ (Fig.~\ref{nonequilibrium_efficiency}(c)). For small $\bar{\mu}$ and $\Delta\mu$, the highest efficiency can appear in both $\Delta > 0$ and $\Delta < 0$ regimes (Figs.~\ref{nonequilibrium_efficiency}(a) and (b)), as particle flows between the charger and battery can either align (both release or absorb particles) or oppose (one releases while the other absorbs).

The results in Fig.~\ref{nonequilibrium_efficiency} also demonstrate how to manipulate nonequilibrium characterized by the chemical potential difference between the reservoirs to enhance the charging efficiency of the quantum battery. When $\bar{\mu}$ is zero or positive, the maximum efficiency is predominantly localized in the regime $\Delta < 0$ under significant chemical potential difference (Figs.~\ref{nonequilibrium_efficiency}(a-c)). However, when $\bar{\mu}$ is negative, such as $\bar{\mu} = -6\lambda$, the role of nonequilibrium becomes negligible for enhancing efficiency.

From the results in Fig.~\ref{nonequilibrium_efficiency}, we observe that the efficiency at resonance ($\Delta = 0$) is always below 50\%. However, this efficiency can exceed 90\% when the charger is driven non-resonantly ($\Delta \neq 0$). Therefore, non-resonant driving proves to be a more effective strategy for enhancing the performance of the quantum battery in our setup.

\begin{figure}
\centering
\includegraphics[width=0.48\columnwidth]{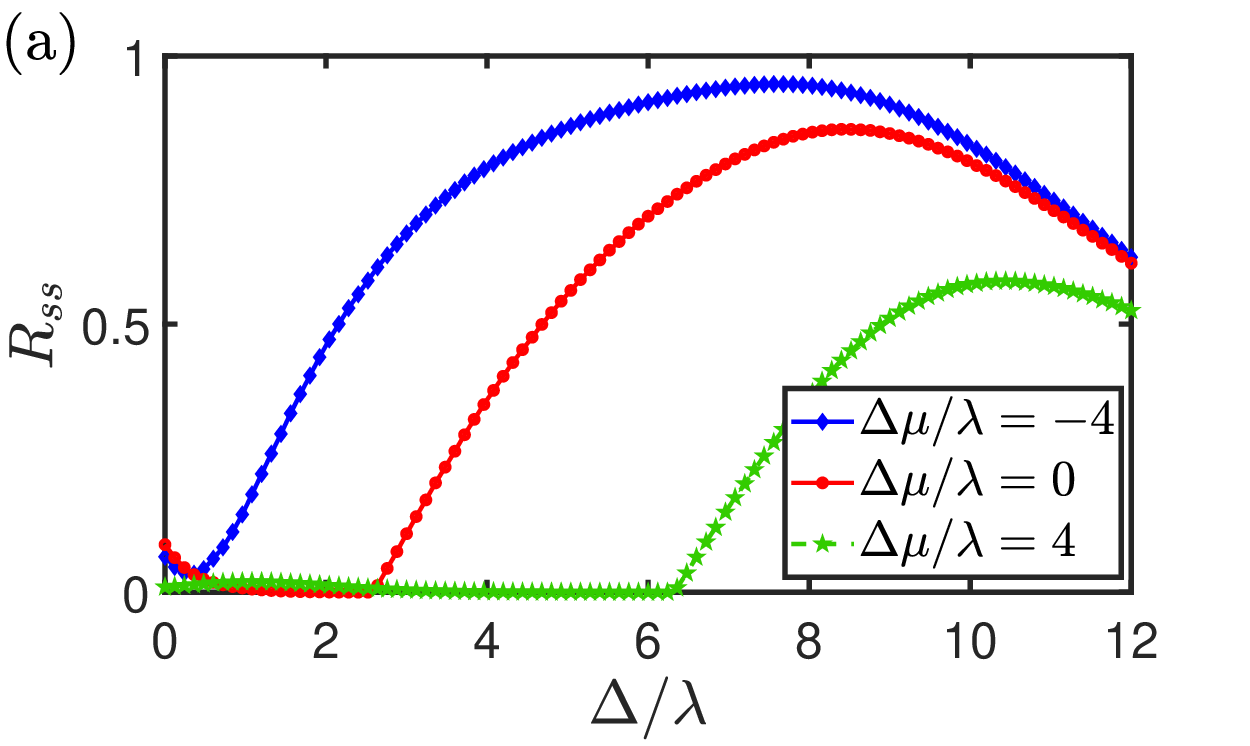}
\includegraphics[width=0.48\columnwidth]{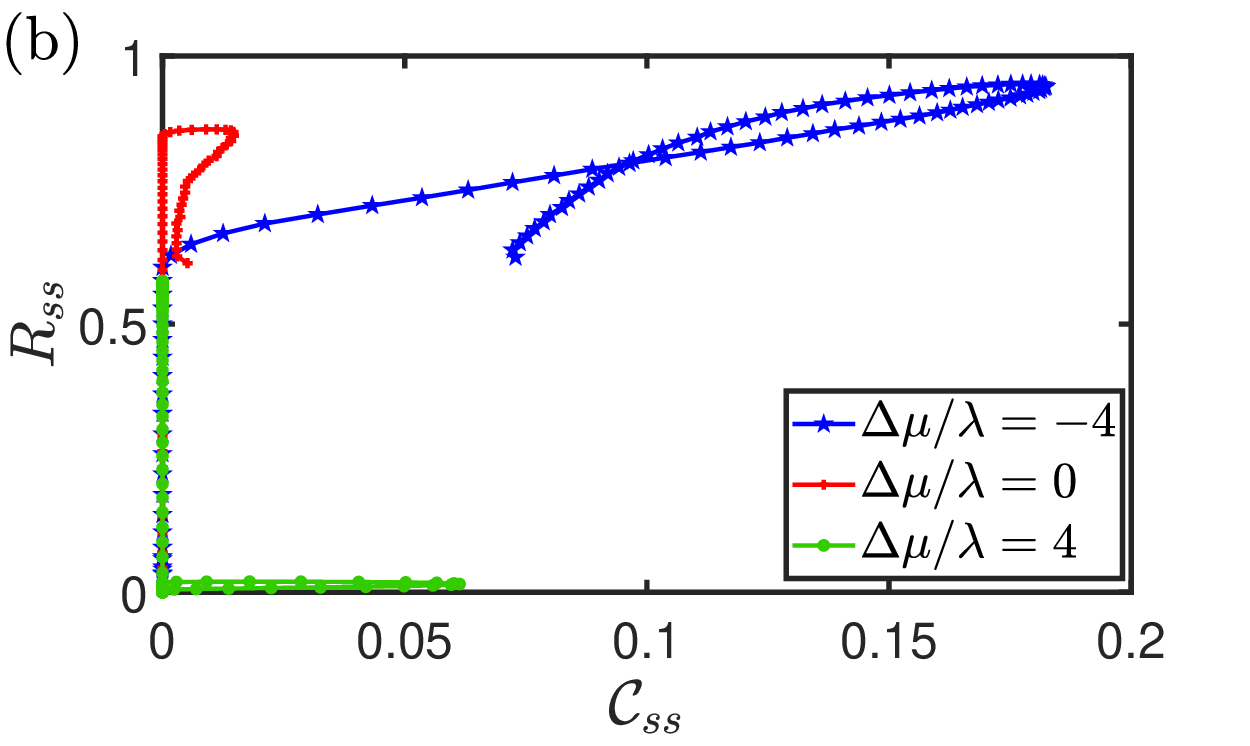}
\caption{The efficiency $R_{\rm ss}$ as a function of the detuning $\Delta$ (a) and the concurrence (b) for different chemical potential difference $\Delta\mu$. The parameters are set as $F = 0.5\lambda$, $\omega_c = 5\lambda$, $T_1 = T_2 = \lambda$, $\bar{\mu} = 2\lambda$, and $\alpha_1 = \alpha_2 = 0.1\lambda$.}
\label{efficiencydelta}
\end{figure}

{At the end of this section, we critically examine the relationship between charger-battery entanglement and charging efficiency in our quantum battery setup. In Fig.~\ref{efficiencydelta}(a), we plot the efficiency as a function of the detuning $\Delta$ for various chemical potential differences $\Delta \mu$, using the same parameters as in Fig.~\ref{concurrence}(a). While both the concurrence in Fig.~\ref{concurrence}(a) and the efficiency in Fig.~\ref{efficiencydelta}(a) display non-monotonic behavior, their line shapes are markedly different. The differences are demonstrated more clearly in Fig.~\ref{efficiencydelta}(b), where the efficiency is plotted against the concurrence. Remarkably, a single concurrence value can correspond to multiple efficiency outcomes, clearly demonstrating that the efficiency is not a monotonic or even uniquely determined-function of the concurrence. This reveals a key finding: contrary to the widely held belief that entanglement inherently enhances quantum device performance, our results show no positive correlation between charger-battery entanglement and charging efficiency. This challenges the conventional assumption and calls for a reevaluation of the role of entanglement in quantum thermodynamic processes.}

\section{Charging power}
\label{power1}
\begin{figure}
\centering
\includegraphics[width=0.48\columnwidth]{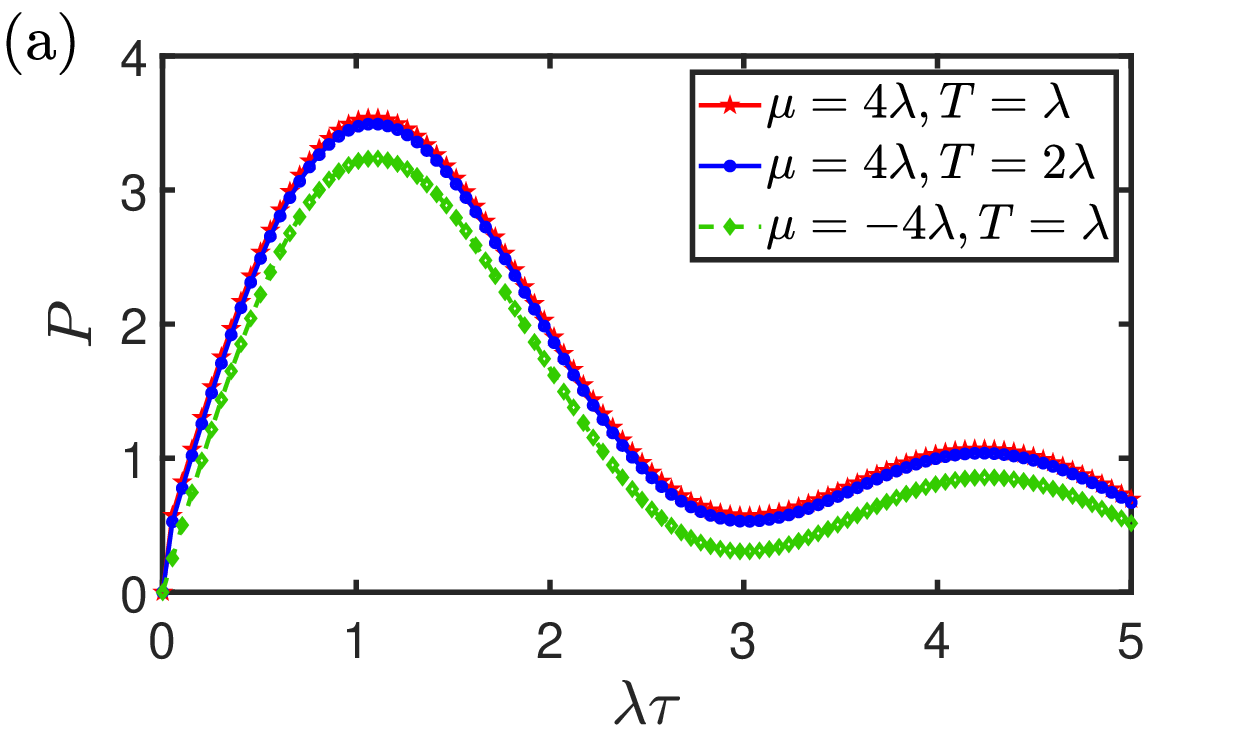}
\includegraphics[width=0.48\columnwidth]{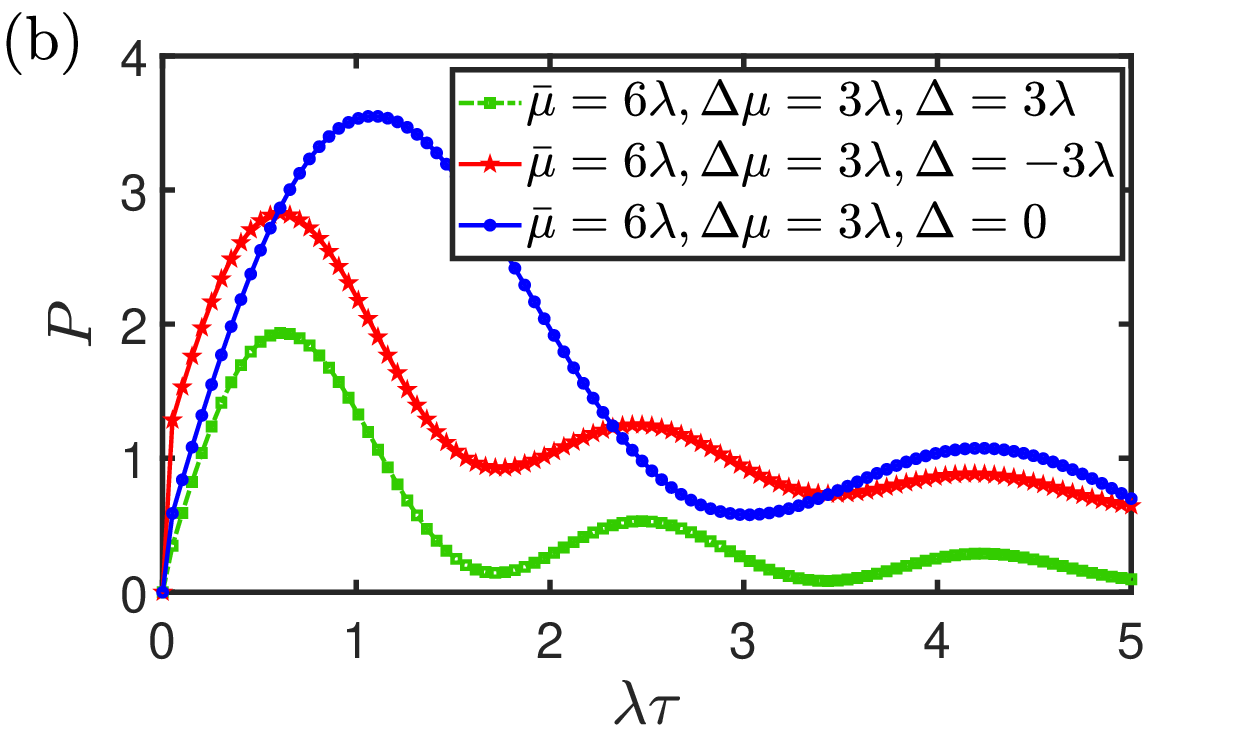}
\caption{The charging power $P(\tau) = E(\tau)/\tau$  in the equilibrium and  nonequilibrium fermion reservoirs. The parameters are set as $\mu_1=\mu_2=\mu, T_1=T_2=T, F=0.5\lambda, \omega_c=5\lambda,\Delta_1=\Delta_2=0,\alpha_1=\alpha_2=0.1\lambda$ for (a) and $T_1=T_2=\lambda, F=0.5\lambda, \omega_c=5\lambda,\bar{\Delta}=0,\alpha_1=\alpha_2=0.1\lambda$ for (b).}
\label{equilibrium_power}
\end{figure}

In the battery scenario, the charging power $P(\tau) = E(\tau)/\tau$ is another key quantity of interest. It describes the ``velocity'' of the charging process, as $E(\tau)$ represents the energy transferred to the battery from the charger. In Figs.~\ref{equilibrium_power}(a) and (b), we analyze the charging power in both equilibrium and nonequilibrium cases.

In the equilibrium case, the results shown in Fig.~\ref{equilibrium_power}(a) indicate that both the chemical potential and the temperature have minimal impact on the charging power. This suggests that an equilibrium system alone cannot effectively enhance the charging power. Fortunately, this limitation can be addressed by transitioning to the nonequilibrium setup.

As illustrated in Fig.~\ref{equilibrium_power}(b), compared to the equilibrium case with $\Delta\mu = 0$, significantly higher charging power can be achieved for $\Delta\mu > 0$ in the regime $\Delta\leq0$. This highlights the pivotal role of the compensation mechanism in enhancing the charging power in a nonequilibrium quantum battery setup.

\section{Conclusion}
\label{con}
In this paper, we have investigated the efficiency of a quantum battery setup where both the charger and the battery are modeled as two-level systems. These systems are coherently coupled to each other and simultaneously interact with their respective fermionic reservoirs, with nonequilibrium characterized by the chemical potential difference. Going beyond the traditional phenomenological master equation, we consider the effects of external driving and charger-battery coupling on the eigen spectrum of the system. By treating these effects in a non-perturbative manner, we derive the Redfield master equation without applying the secular approximation. Solving this master equation, we reveal the significant role of nonequilibrium reservoirs in the performance of the quantum battery setup. Generally, the Redfield master equation does not guarantee the positivity of the density matrix, as the eigen values of the density matrix may become negative. However, we have verified that the density matrix remains positive throughout the entire parameter regime considered in this study. Our results demonstrate that when the charger is driven non-resonantly, the chemical potential difference in the fermionic reservoirs can enhance the quantum battery's efficiency via a compensation mechanism. This mechanism is particularly effective under non-resonant driving and can also be employed to boost the charging power of the quantum battery under significant nonequilibrium conditions. {In addition, we challenge the long time belief that the entanglement can enhance the performance of quantum device by showing that there is no positive correlation between the charger-battery entanglement and charging efficiency.}  In summary, nonequilibrium reservoirs provide an effective framework for designing energy devices based on open quantum systems.

\section*{Author contributions}

Z. Wang and J. Wang designed the research. Z. Wang and H. Yu performed the main calculations. Z. Wang and J. Wang wrote the manuscript. All of the authors contributed to the planning and made revision of the manuscript.

\section*{Acknowledgments}

Science and Technology Development Project of Jilin Province (Grant No. 20230101357JC) and  National Natural Science Foundation of China (Grants No. 12375010, 21721003 and 12234019)

\end{document}